\documentclass[11pt,a4paper]{article}
\usepackage{jcappub}
\usepackage{graphicx}	
\usepackage{amsmath, cases}
\usepackage{caption}
\usepackage{subcaption}
\usepackage{xcolor}

\title{Charged Binaries in Gravitational Tides}

\author[a,b]{Elisa Grilli,}
\author[a,b]{Marta Orselli,}
\author[b]{David Pereñiguez}
\author[a,b]{\\ and Daniele Pica}

\affiliation[a]{Dipartimento di Fisica e Geologia, Universit\`a di Perugia,
\\
I.N.F.N. Sezione di Perugia,\\
Via Pascoli, I-06123 Perugia, Italy}

\affiliation[b]{Niels Bohr International Academy, Niels Bohr Institute, Copenhagen University,\\
Blegdamsvej 17, DK-2100 Copenhagen, Denmark}

\emailAdd{elisa.grilli@nbi.ku.dk}
\emailAdd{orselli@nbi.dk}
\emailAdd{david.pereniguez@nbi.ku.dk}
\emailAdd{daniele.pica@nbi.ku.dk}

\abstract{Next-generation low-frequency interferometers are expected to detect binary systems near supermassive black holes, where tidal effects can alter significantly the binary's motion. This motivates a broader investigation of how external gravitational fields influence the dynamics of physical systems. In this work, we consider a charged black hole binary system subject to a gravitational tide. We first construct a stationary gravitational tide acting on a dyonic Reissner–Nordström black hole and, focusing on the extreme mass-ratio limit, we analyze the motion of a test particle. By calculating the particle's secular Hamiltonian, we obtain the ISCO and light ring tidal shifts in terms of explicit functions of the binary's parameters. Our results show that tidal corrections are suppressed as the black hole’s charge increases, but they persist in the extremal limit yielding a finite contribution. This work paves the way towards studying tidal effects on other charged systems, such as topological stars.}

\begin{document}
\maketitle
\clearpage
\section{Introduction} \label{Sec:introduction}

The strong-field regime of gravitation has remained both fascinating and elusive for decades. Its most fundamental prediction, black holes, have challenged our deepest ideas about the structure of spacetime and remain at the core of gravitational physics since their theoretical discovery. The fact that gravitational wave astronomy provides a direct channel of observation into such regime of gravity is, consequently, an important milestone in science and motivates studying the effect of strong gravitational fields on physical systems. 

In particular, understanding the response of self-gravitating systems to slowly-varying tidal fields has proved being very valuable. On the one hand, the deformability properties of stars under gravitational tides reveal information about their structure and can be used, for example, to constrain the equation of state of nuclear matter \cite{Flanagan:2007ix,Hinderer:2007mb,Hinderer:2009ca}. On the other hand, quite intriguingly black holes exhibit no response to external tides,\footnote{In fact, this is true not only for gravitational tides but also for  other kinds of fields, like scalar or electromagnetic ones. More exotic environments or curvature corrections can, however, yield nonvanishing tidal responses \cite{Cardoso:2019upw,Katagiri:2024fpn}.} a fact that is made somehow precise by showing that their tidal Love numbers vanish \cite{Binnington:2009bb}. Given that in a binary merger tidal deformations can leave imprints in the last stages of the inspiral phase, this can be used to learn about the nature of the binary's constituents. 

Another paradigmatic way of understanding tidal deformations is via the motion of test fields (say, a test particle) in the vicinity of the deformed object. This is, in fact, a case of astrophysical relevance as next generation interferometers are expected to observe black hole binary systems next to supermassive black holes (SMBHs) \cite{Zhang:2024ibf, Chen:2018axp, cite-key, PhysRevD.106.103040}. An interesting case is when the binary itself has one component much larger than the other, thus yielding a hierarchical triple. These type of systems have been intensively studied in the literature in the Post Newtonian (PN) approximation, where each of the three objects is treated as a test particle, but still keeping the hierarchical regime~\cite{Amaro_Seoane_2018, Amaro_Seoane_2021, Amaro_Seoane_2007, berry2019uniquepotentialextrememassratio}. However, through processes like the migration trap mechanism, the binary systems may be placed only a few Schwarzschild radii away from the SMBH~\cite{Peng:2021vzr, Bellovary_2016, Secunda_2021}, in which case the PN expansion ceases to hold. In some sense, these systems generalise Extreme Mass Ratio Inspirals (EMRIs) by endowing the secondary with structure (it gets replaced by another EMRI), and accounting for strong field effects requires a computation in full General Relativity (GR). A first step towards understanding the effect of strong external fields on EMRIs was performed in~\cite{Yang:2017aht, Camilloni:2023rra, Cardoso_2021} where, in particular, the smallest object of the triple behaves like a test particle evolving on a tidally deformed black hole. Another interesting situation is the one examined in \cite{Camilloni:2023xvf} where it is shown how strong gravitational effects modify the dynamics of a binary system of  black holes with comparable masses.

In general, it is important to understand from a theoretical point of view the behaviour of self-gravitating systems when subject to external tides. While Love numbers have been studied for a wide variety of objects and theories, the behaviour of matter surrounding the deformed object has remained less explored. In this paper we consider the motion of test particles in the vicinity of a tidally deformed black hole carrying both electric and magnetic charge (a dyonic black hole). In particular, we analyse the dependence on the hole's charge of the tidal deformations to the Innermost Stable Circular Orbit (ISCO) and the light ring.

Being well defined both physically and mathematically, charged black holes have been largely explored in the literature of theoretical astrophysics for decades and yielded very valuable lessons \cite{Zerilli:1974ai,Johnston:1974vf,PhysRevD.10.1057,Gerlach:1979rw,Gerlach:1980tx,Chandrasekhar1979OnTM}. In fact, while typically black holes are expected to quickly neutralise by either friction with interstellar medium or Schwinger pair-decay, there are physical mechanisms that allow them to retain their charge (although small) \cite{Wald:1974np}. Assuming they are near-extremal, charged black holes could constitute a significant fraction of dark matter \cite{Kritos:2021nsf} or, alternatively, their charge could be due to  minicharged dark matter, hidden vector fields \cite{DeRujula:1989fe,Perl:1997nd,Holdom:1985ag,Sigurdson:2004zp,Davidson:2000hf,McDermott:2010pa,Cardoso:2016olt,Khalil:2018aaj,Bai:2019zcd,Gupta:2021rod} or magnetic monopoles \cite{Polyakov:1974ek,tHooft:1974kcl,Maldacena:2020skw,Gibbons:1976sm}. In the latter case one has magnetic black holes, which have gained a lot of attention recently. When interacting with charged matter they behave drastically different than their electric counterparts, and this can be seen at several levels including the binary dynamics \cite{Liu:2020vsy,Liu:2020bag,Chen:2022qvg,Pereniguez:2023wxf}, black hole superradiance \cite{Pereniguez:2024fkn}, hairy black hole solutions \cite{Gervalle:2024yxj,Cunha:2024gke}, and other physical processes (see \cite{Dyson:2023ujk} and references therein).\footnote{For magnetic black holes in some dark matter models see \cite{DeFelice:2024eoj}.} In addition, some well-defined models of horizonless compact objects, such as topological stars \cite{Bah:2020ogh} (see also \cite{Dima:2024cok,Bena:2024hoh}), exhibit charges of magnetic type. These are interesting, yet more complicated, solutions so solving the black hole case first is an instructive example. Finally, given that charged black holes do also share some similarities with rotating ones, such as the existence of an extremal limit, the results in the charged case may anticipate the behaviour in the more complicated scenario where the black hole carries angular momentum.

This paper is organised as follows. 
In Sec.~\ref{Sec:Tides} we construct the metric for a tidally deformed dyonic Reissner--Nordström black hole. The leading order (quadrupolar) solution is then used in Sec.~\ref{Dynamics} to study the dynamics of a test particle moving in such a background. In particular, we focus on the secular effects induced by the tidal deformation and we derive the secular Hamiltonian for the test particle. As an application of our approach, in Sec.~\ref{ISCOstuff} we determine how the tidal deformation modifies the location and properties of the ISCO (for neutral and charged particles) and of the light ring. Finally Sec.~\ref{Sec:Discussion} contains our conclusions.

\section{Tides on Dyonic Reissner--Nordström Black Holes} \label{Sec:Tides}

\begin{figure*}
    \centering
    \includegraphics[width=0.50\textwidth]{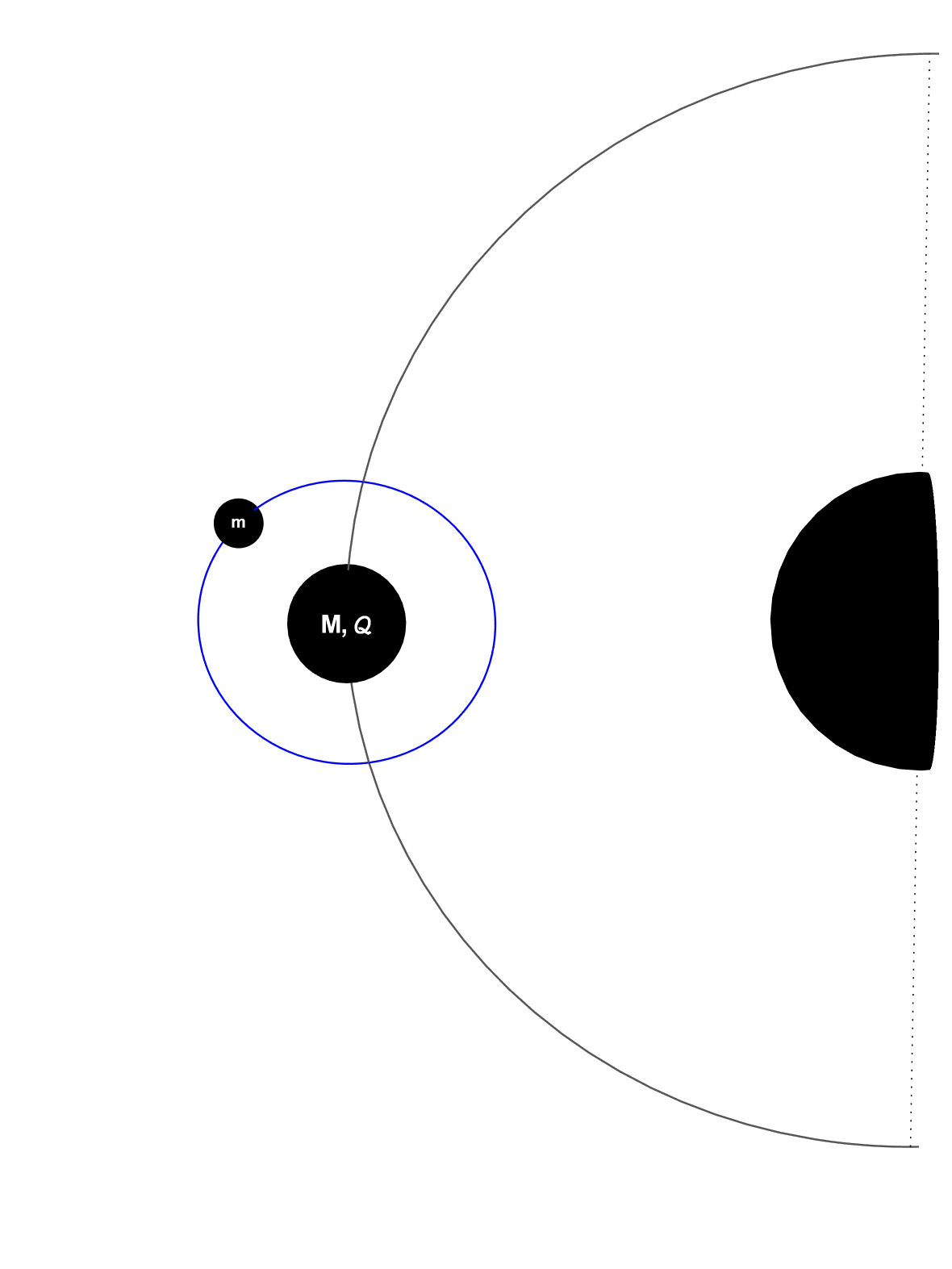}
    \caption{A hierarchical triple system, where the binary composed by a charged RN black hole (with mass $M$ and total charge $\mathcal{Q}$) and a test particle (with mass $m$) is subject to the gravitational tide created by a larger neutral black hole.}
    \label{fig:triple}
\end{figure*}

 Our aim is to describe the dynamics of a small body evolving in the neighbourhood of a dyonic black hole, when the system is subject to an external gravitational tide (e.g. that created by a much larger Kerr black hole, see Fig.~\ref{fig:triple}). This can be done effectively by considering the motion of a structureless (yet possibly charged) test particle in the background of a dyonic black hole which is distorted by a tidal field. The goal of this section is to construct such a background.

\subsection{The Dyonic Reissner--Nordström Black Hole}

We will consider the Einsiten--Maxwell theory, where the most general spherically-symmetric and asymptotically flat black hole solution is the dyonic Reissner--Nordström (RN) spacetime. In geometric units $G=c=1$, the solution reads
\begin{equation}
\begin{aligned}\label{RN}
ds^{2}&=-f(r)dt^{2}+\frac{dr^{2}}{f(r)}+r^{2}d\Omega^{2}\, , \quad f(r)=1-\frac{2M}{r}+\frac{\mathcal{Q}^2}{r^{2}}\, , \\
\mathcal{F}&=-\frac{Q}{r^{2}}dt\wedge dr+P\epsilon\, .
\end{aligned}
\end{equation}
Here $M$ is the mass of the black hole, $d\Omega^{2}$ and $\epsilon$ are, respectively, the metric and volume form of the round 2-sphere. On the latter, it will be convenient to work both with generic and spherical coordinates, $z^{A}$ (where $A=1,2$) and $(\theta,\phi)$ respectively, so that one has
\begin{equation}
    d\Omega^{2}=\Omega_{AB}dz^{A}dz^{B}=d\theta^{2}+\sin^{2}\theta d\phi^{2}\, , \quad \epsilon=\frac{1}{2!}\epsilon_{AB}dz^{A}\wedge dz^{B}=\sin\theta d\theta\wedge d\phi\, ,
\end{equation}
and we may often use $D_{A}$ to denote the covariant derivative on the 2-sphere. In \eqref{RN}, the electric charge $Q$, magnetic charge $P$, and ``total'' charge $\mathcal{Q}$ are defined by
\begin{equation}
    Q=\frac{1}{4\pi}\int_{S^{2}} \star F\,, \ \ P=\frac{1}{4\pi}\int_{S^{2}} F\,, \ \ \mathcal{Q}=\sqrt{Q^2+P^2}\, ,
\end{equation}
where $S^{2}$ is any 2-sphere enclosing the hole. There are Killing horizons at 
\begin{equation}
\label{eq:eh_rn}
    r_{\pm}=M\pm\sqrt{M^{2}-Q^{2}-P^{2}}
\end{equation}
as long as $M^{2}\geq Q^{2}+P^{2}$, where $r_{+}$ and $r_{-}$ are event and Cauchy horizons, respectively. Given that the coordinates $(t,r,\theta,\phi)$ are singular at $r=r_{\pm}$, it will be useful to work with advanced Eddington--Finkelstein (EF) coordinates $(v,r,\theta,\phi)$, where the new coordinate is defined as
\begin{equation}
    dv=dt+\frac{dr}{f(r)}\,.
\end{equation}
These coordinates are well defined at the future event horizon $r_{+}$ and the metric is written in the \textit{light-cone gauge}, as defined in \cite{Poisson:2009qj}, which we shall follow in the next sections.

\subsection{Stationary Gravitational Tides}

In the presence of a stationary tidal source, the dyonic RN spacetime is deformed to
\begin{equation}
\label{tdmetric}
    g_{\mu\nu}=g^{\text{dRN}}_{\mu\nu}+h_{\mu\nu}\, , \quad \mathcal{F}_{\mu\nu}=\mathcal{F}^{\text{dRN}}_{\mu\nu}+\delta \mathcal{F}_{\mu\nu}\, ,
\end{equation}
where $(g^{\text{dRN}}_{\mu\nu},\mathcal{F}^{\text{dRN}}_{\mu\nu})$ is the dyonic RN solution, and $(h_{\mu\nu},\delta \mathcal{F}_{\mu\nu})$ is a stationary solution to the linearised Einstein--Maxwell equations.\footnote{Notice that, since the background is charged, even if the tidal source is electromagnetically neutral it will excite $\delta \mathcal{F}_{\mu\nu}$ necessarily.} Tidal fields on electrically charged RN spacetimes have been intensively studied in the literature in several contexts. In particular, the response to static tides by electric black holes, encoded in the tidal Love numbers and electromagnetic polarisabily, was studied in \cite{Cardoso:2017cfl,Poisson:2021yau} and recently in \cite{Rai:2024lho}, while the generalisation to higher dimensional spacetimes was considered in \cite{Pereniguez:2021xcj} and later on in \cite{Charalambous:2024tdj}. 
\\ \\
Here, we focus on four spacetime dimensions and construct the most general (yet linear) gravitational, stationary tidal field in the neighbourhood of a dyonic RN spacetime. Using the formalism devised in \cite{Pereniguez:2023wxf}, constructing the solution is straightforward, so we will spare the reader of the details and simply present the result.\footnote{Note that since the equations of motion of the Einstein--Maxwell theory are invariant under electric-magnetic duality transformations, an alternative approach consists in taking purely electric solutions 
obtained using the results of \cite{Cardoso:2017cfl,Poisson:2021yau,Rai:2024lho} and duality-rotate them into a general dyonic solution.} In Schwarzschild coordinates and in the Regge--Wheeler gauge, a general gravitational tidal field on a dyonic RN spacetime is given by
\begin{align}\label{h} \notag
    h&=\mathcal{X}_{\ell m}(r)Y^{\ell m}\left(dt\otimes dt+\frac{dr\otimes dr}{f(r)^2}\right)+r^{2}\mathcal{U}_{\ell m}(r)Y^{\ell m}d\Omega^{2}\\ 
    &+r^{2}\mathcal{V}_{\ell m}(r)\left(dt \otimes X^{\ell m}+ X^{\ell m}\otimes dt \right)\, , \\ \notag \\ \notag
    \delta \mathcal{F} &= \left(Q \frac{\mathcal{U}_{\ell m}(r)}{r^{2}}-\frac{\mathcal{E}_{\ell m}(r)}{r^{2}}\right)Y^{\ell m}dt\wedge dr+\left(P\mathcal{V}_{\ell m}(r)-\frac{f(r)}{\ell(\ell+1)}\frac{d\mathcal{E}_{\ell m}(r)}{dr}\right)dt \wedge Z^{\ell m}\\ \label{F}
    &+\frac{f(r)}{\ell(\ell+1)}\frac{d\mathcal{B}_{\ell m}(r)}{dr}dr\wedge X^{\ell m}+ \mathcal{B}_{\ell m}(r)Y^{\ell m}\epsilon
    \, .
\end{align}
\\ 
The solution is given in terms of the usual (complex) spherical harmonics $Y^{\ell m}$, together with the associated even and odd vector harmonics $Z^{\ell m}_{A}=D_{A}Y^{\ell m}$ and $X^{\ell m}_{A}=\epsilon_{AB}D^{B}Y^{\ell m}$ (see e.g.\cite{Pereniguez:2023wxf}). We have introduced the functions\footnote{For convenience, here $\Phi^{+}$ and $\Phi^{-}$ are $1/(2\mathcal{Q}^{2})$ and $i/(2\mathcal{Q}^{2})$ times the definitions of $\Phi^{\pm}$ in \cite{Pereniguez:2023wxf}, respectively.}
\begin{equation}
\mathcal{E}_{\ell m}(r)\equiv Q \Phi^{+}_{\ell m}(r)+ P\Phi^{-}_{\ell m}(r)\, , \quad \mathcal{B}_{\ell m}(r)\equiv P \Phi^{+}_{\ell m}(r)- Q\Phi^{-}_{\ell m}(r)\, ,
\end{equation}
so the solution \eqref{h}-\eqref{F} depends on the radial functions $\mathcal{X}_{\ell m}(r),\mathcal{U}_{\ell m}(r),\mathcal{V}_{\ell m}(r)$ and $\Phi^{\pm}_{\ell m}(r)$. Remarkably, these turn out to be finite power series in $r$ of the forms
\begin{align}\label{solEVEN}
\mathcal{X}_{\ell m}(r)&= \ r^{\ell}\sum_{n=0}^{\ell+3}a^{\ell}_{n}r^{-n} \,, \ \ \mathcal{U}_{\ell m}(r)= \ r^{\ell}\sum_{n=0}^{\ell+1}b^{\ell}_{n}r^{-n} \,, \ \ \Phi^{+}_{\ell m}(r)= \ r^{\ell}\sum_{n=0}^{\ell}c^{\ell}_{n}r^{-n}  \ \, ,\\ \label{solODD}
\mathcal{V}_{\ell m}(r)&= \ r^{\ell-1}\sum_{n=0}^{2\ell}d^{\ell}_{n}r^{-n} \,, \quad \Phi^{-}_{\ell m}(r)= \ r^{\ell}\sum_{n=0}^{\ell+1}e^{\ell}_{n}r^{-n}\, ,
\end{align}
where the constant coefficients $a^{\ell}_{n},b^{\ell}_{n},c^{\ell}_{n},d^{\ell}_{n},e^{\ell}_{n}$ can be obtained immediately from the equations of motion (we exhibit explicitly a few of them in Appendix \ref{AppExpSol}, as an example). 
\\ \\
A few comments are in order at this point. First, the functions $\{\mathcal{V}_{\ell m}(r),\Phi^{-}_{\ell m}(r)\}$ and $\{\mathcal{X}_{\ell m}(r),\mathcal{U}_{\ell m}(r),\Phi^{+}_{\ell m}(r)\}$ are subject to two decoupled sets of equations, and thus are independent from each other. In the terminology of \cite{Pereniguez:2023wxf}, the former set of functions belongs to the generalised odd sector of the fluctuation, while the latter belongs to the generalised even one.\footnote{Notice that if the magnetic black hole charge is nonvanishing, $P\ne0$, the generalised even and odd sectors mix the traditional even and odd pieces of the fluctuation \cite{Zerilli:1974ai}.} Second, the boundary conditions chosen in deriving the solution \eqref{h}-\eqref{F} are as follows. At the horizon one of course requires regularity of the fluctuation. At infinity one has that in general the leading behaviour of the electromagnetic master variables is $\Phi^{\pm}_{\ell m}\sim r^{\ell+1}$. Thus, in order to match our field to a purely gravitational tide (instead of a gravitational and electromagnetic one), we impose that $\Phi^{\pm}_{\ell m}$ grow slower instead, as $\sim r^{\ell}$ (for a similar discussion see \cite{Cardoso:2017cfl,Pereniguez:2021xcj,Rai:2024lho}). These boundary conditions determine uniquely the solutions for the generalised even and odd sectors (up to a global amplitude per sector, which should be matched to the distant source of the tidal field). In particular, we notice that our $\delta \mathcal{F}$ vanishes as the black hole charges $P$ and $Q$ go to zero, as expected for a purely gravitational tide. If there was an electromagnetic tide too, then a part of $\delta \mathcal{F}$ would survive as $P$ and $Q$ vanish. Similarly, since dipolar ($\ell=1$) degrees of freedom are entirely due to the electromagnetic field, and we are not considering electromagnetic tidal sources, there are no dipolar modes in our solution. It is also worth noting that a corollary of the solution \eqref{h}-\eqref{F} is that the tidal Love numbers of dyonic RN black holes vanish, since the unique solution which is regular at the event horizon and matches a gravitational tide at infinity does not include a response of internal multipoles.

Finally, for the subsequent analysis it will be enough to restrict to a purely electric black hole, and we will focus our attention on the quadrupolar modes $\ell=2$. In order to write our solution in terms of the tidal multipole moments, we will follow \cite{Poisson:2009qj} and write \eqref{h}-\eqref{F} in the light-cone gauge and horizon locking coordinates. This is easily done using the gauge-transformation rules in \cite{Pereniguez:2023wxf}, and one finds that the non-vanishing components of the fluctuation read
\begin{equation}
\begin{aligned}\label{hPVgauge}
h_{vv}&= 
-r^2 f^{2}\mathcal{E}^{\mathsf{q}}\,,\\
h_{vA}&=
-\frac{2}{3}r^{3}f\left(\mathcal{E}^{\mathsf{q}}_{A}-\mathcal{B}^{\mathsf{q}}_{A}\right)\, ,\\
h_{AB}&=-\frac{1}{3} r^2  \left(Q^2+r^2-2 M^2\right)\mathcal{E}^{\mathsf{q}}_{AB}+\frac{r^{2}}{3}\left(r^{2}-r^{2}_{+}\right)\mathcal{B}^{\mathsf{q}}_{AB}\, ,
\end{aligned}    
\end{equation}
and
\begin{equation}\label{FPVgauge}
\begin{aligned}
\delta \mathcal{F}_{vr}&= \frac{Q}{2 r^{2}}(r^{2}-Q^{2})\mathcal{E}^{\mathsf{q}}\,,\\
\delta\mathcal{F}_{vA}&=Q r f \mathcal{E}^{\mathsf{q}}_{A}\, ,\\
\delta\mathcal{F}_{rA}&=-\frac{r}{3}Q  \left(\mathcal{E}^{\mathsf{q}}_{A}-\mathcal{B}^{\mathsf{q}}_{A}\right)\, ,\\
\delta \mathcal{F}_{AB}&=\frac{Q}{2}\left(r^{2}-Q^{2}\right)\mathcal{B}^{\mathsf{q}}\epsilon_{AB}\, ,
\end{aligned}    
\end{equation}
where $\mathcal{E}^{\mathsf{q}},\mathcal{E}^{\mathsf{q}}_{A},\mathcal{E}^{\mathsf{q}}_{AB}$ and $\mathcal{B}^{\mathsf{q}},\mathcal{B}^{\mathsf{q}}_{A},\mathcal{B}^{\mathsf{q}}_{AB}$ are the electric and magnetic tidal multipole potentials as introduced in \cite{Poisson:2009qj}. The solution \eqref{hPVgauge}-\eqref{FPVgauge} is the background we will employ next to study the test particle dynamics.

\section{Secular Dynamics}
\label{Dynamics}
The dynamics of a particle of mass $m$ in the tidally deformed dyonic RN spacetime is described by the geodesic equation, taking into account the corrections to the dyonic RN geometry up to linear order (see \eqref{tdmetric}-\eqref{F}). This is an accurate approximation as long as $m/M$ is small enough, so that self-force corrections are negligible. In addition, we also require that $M/\mathcal{R}\ll 1$, where  $\mathcal{R}$ is the length-scale of the external tidal field, as in the \textit{small-tide approximation}~\cite{Poisson:2009qj,Camilloni:2023rra}. Expanding in $M/\mathcal{R}$, it can be seen that the quadrupolar multipole ($\ell=2$) yields the dominant effect, so we can restrict to that tidal mode and neglect higher multipoles ($\ell>2$). In addition, since we will be considering only the motion of a test (charged) particle, the equations possess electric-magnetic duality and we are free to choose the duality frame where the black hole only carries electric charge.\footnote{If we were to consider, say, a minimally-coupled charged scalar field then electric-magnetic duality would no more be a symmetry and a purely electric black hole would not be the most general case, see \cite{Pereniguez:2023wxf,Pereniguez:2024fkn}. In that situation, one would need to work with the general solution \eqref{h}-\eqref{F}.} Under these assumptions, the tidal field is given by \eqref{hPVgauge}-\eqref{FPVgauge}.

\subsection{Secular Hamiltonian of the Test Particle }

We consider first the case of a neutral particle evolving on our tidally deformed background. Its four-velocity can be written as 
\begin{equation}
    u^{\mu}=\bar{u}^{\mu}+u_{(1)}^{\mu}\, ,
\end{equation}
where $\bar{u}^{\mu}$ is tangent to a geodesic of the undeformed background, and $u_{(1)}^{\mu}$ is a correction due to the tidal deformation $\sim h_{\mu\nu}$. The Hamiltonian governing the particle dynamics including the leading-order tidal corrections is (we raise and lower indices with the background 
metric \eqref{RN})
\begin{equation}\label{hamiltonian}
    H\,=  \frac{1}{2} \Bar{u}_\mu \Bar{u}^\mu+\Bar{u}_{\mu} u^{\mu}_{(1)}+\frac{1}{2} \Bar{u}^\mu \Bar{u}^\nu h_{\mu\nu}\,.
\end{equation}
We will focus on quasi-circular trajectories, which are small deviations with respect to the background circular orbits given by  $\Bar{u}^\mu=\bigl(\frac{\Bar{E}}{f},0,0,\frac{\Bar{L}}{r^2}\bigr)$, where the constants $\Bar{E}$ and $\Bar{L}$ denote the specific energy and angular momentum. In the presence of the tidal field, the deformed trajectory consists of a mean circular orbit and an oscillatory piece with small amplitude and frequency of order $\sim 1/M$, where $M$ is the black hole's mass. A powerful simplification consists in integrating out the oscillatory piece, and reduce the problem to an effective theory for the mean circular orbit. This captures the ``secular'' effects induced by the tidal deformation (we refer the reader to \cite{Yang:2017aht,Camilloni:2023rra} for more details on the effective description of quasi-circular trajectories). Denoting by $\gamma$ the mean circular orbit, the \textit{secular average of a quantity $\mathcal{A}$} is defined as
\begin{equation}
	\left<\mathcal{A}\right> \equiv \frac{1}{2\pi} \int_{0}^{2\pi} \mathcal{A} \left. \right|_\gamma d\phi \, ,
\end{equation}
and taking the average of the Hamiltonian \eqref{hamiltonian} one finds
\begin{equation}\label{averageH}
    \left \langle H \right \rangle = -\frac{1}{2} \Biggl( \frac{\left \langle E \right \rangle ^2}{f}-\frac{\left \langle L \right \rangle^2}{r^2} \Biggr)+\frac{r^{2}}{2}\left \langle \mathcal{E}^{\mathsf{q}} \right \rangle  \Biggl[ \left \langle E \right \rangle^2\, + f \frac{\left \langle L \right \rangle^2}{r^2} \Biggr]\, .
\end{equation}
Here, $E\equiv-u\cdot\partial_{t}$ and $L\equiv u\cdot \partial_{\phi}$ denote the specific energy and angular momentum of the trajectory and we used that, form the solution \eqref{hPVgauge}, the metric components average to
\begin{equation}
\label{eq:tidal metric components}
   \left \langle h_{vv} \right \rangle\, =\, -r^2\, f^2 \left \langle \mathcal{E}^{\mathsf{q}} \right \rangle,~~~~
    \left \langle h_{v\phi} \right \rangle\, =0 ,~~~~ \left \langle h_{\phi \phi} \right \rangle\, =-r^4 f \left \langle \mathcal{E}^{\mathsf{q}} \right \rangle. 
\end{equation}
Using the averaged or secular Hamiltonian \eqref{hamiltonian} one can study in a simple manner the secular tidal effects induced on the innermost stable circular orbit (ISCO) and the light ring. These are important notions since they encode valuable information about the propagation of matter, light and gravity in the vicinity of the black hole. We study them in the next section.

\section{Tidal Effects on the ISCO and the Light Ring} \label{ISCOstuff}

In this section we focus our attention on the ISCO and the light ring of a charged black hole and compute the shifts induced by an external tidal field on the parameters characterizing these two orbits. We provide analytical expressions in terms of the mass and charge of the RN black hole, both for the unperturbed parameters as well as for their tidal correction, for the case of a neutral and charged test particle.

\subsection{The Case of a Neutral Particle}

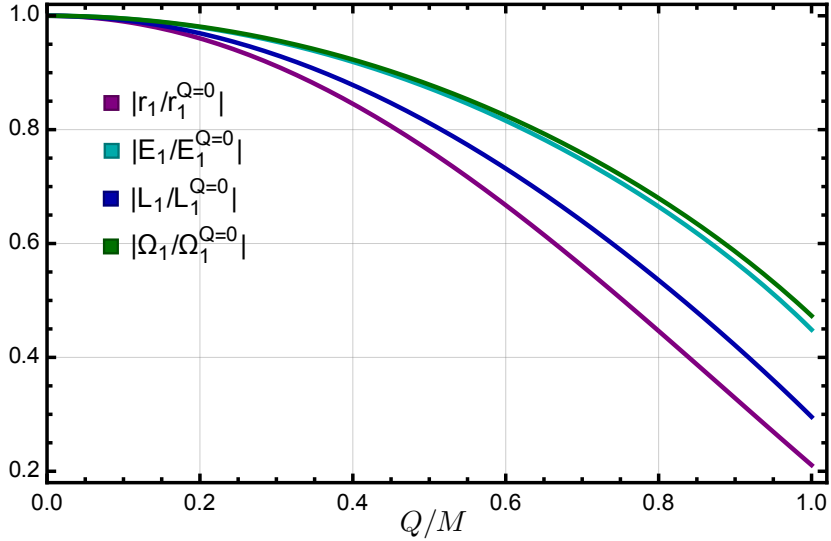
\begin{figure*}
    \centering
    \includegraphics[width=0.70\textwidth]{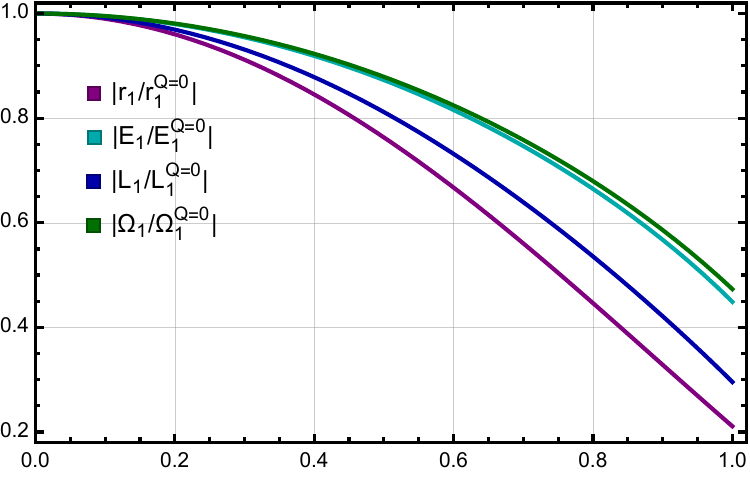}
    \begin{picture}(0,0)
        \put(-165,-4){$Q/M$ }
    \end{picture}
    \caption{Tidal corrections to the ISCO's position, energy, angular momentum and orbital frequency as a function of the black hole's charge-to-mass ratio $Q/M$. We represent the absolute value of the tidal correction normalised to their value in the case that the black hole is neutral.}
    \label{unc_corrections}
\end{figure*}

Following the reasoning of \cite{PhysRevLett.113.161101,Camilloni:2023rra}, the position of the ISCO, $r_{\text{ISCO}}$, is determined by the following conditions in terms of the secular Hamiltonian \eqref{averageH},
\begin{equation}
\label{ISCO_cond}
    \langle H\rangle \vert_{r=r_{\rm  ISCO}}=-\frac{1}{2}~,
    ~~
    \frac{d\langle H\rangle}{dr}\bigg\vert_{r=r_{\rm ISCO}}=0~,
    ~~
    \frac{\partial^2 \langle H\rangle }{\partial r^2}\bigg\vert_{r=r_{\rm ISCO}}=0~.
\end{equation}
Its position should be a deviation, from the unperturbed ISCO, proportional to the mean amplitude of the tidal field $\sim \langle\mathcal{E}^{\mathsf{q}}\rangle$, and similarly for the energy and angular momentum $E_{\rm ISCO}$ and $L_{\rm ISCO}$. Up to first order in $\langle\mathcal{E}^{\mathsf{q}}\rangle$, we can write them as
\begin{equation}
\label{parameters_exp}
    r_{\rm ISCO}\simeq r_0-\frac{M^{2}}{2}\langle\mathcal{E}^{\mathsf{q}}\rangle\, r_1 \,,~~~~
    L_{\rm ISCO}\simeq L_0-\frac{M^{2}}{2}\langle\mathcal{E}^{\mathsf{q}}\rangle\, L_1\,,~~~~
    E_{\rm ISCO}\simeq E_0-\frac{M^{2}}{2}\langle\mathcal{E}^{\mathsf{q}}\rangle\, E_1\,,
\end{equation}
where $\left(r_0, L_0, E_0\right)$ denote the unperturbed ISCO parameters, and attempt to solve \eqref{ISCO_cond} for the deformations $\left(r_1, E_1, L_1\right)$. For completeness we will also consider the ISCO's orbital frequency, which is given by \cite{Camilloni:2023rra, Detweiler:2008ft,Yang:2017aht, Cardoso_2021}
\begin{equation}
    \label{Omega_def}
    \Omega^2=\left(\frac{u^{\phi}}{u^{t}}\right)^2=\frac{1}{2r^2}\left[\frac{2M}{r}-\left(r-3 M\right)u^\mu u^\nu\partial_r \langle h_{\mu\nu}\rangle\right]~,
\end{equation}
and search for solutions of the form 
\begin{equation}
    \label{omega_exp}
    \Omega_{\text{ISCO}}\simeq \Omega_0 -\frac{M^{2}}{2}\langle\mathcal{E}^{\mathsf{q}}\rangle \ \Omega_1\, .
\end{equation}
The ISCO parameters in the absence of a tidal deformation can be readily obtained by solving \eqref{ISCO_cond} at zeroth order in $\langle\mathcal{E}^{\mathsf{q}}\rangle$. In terms of the black hole's mass and charge, we have 
\begin{equation}\label{eq:unp_uncharged} 
\begin{aligned}
    r_0&= 2M + \left(\frac{\mathcal{W}}{M}\right)^{1/3}+\left(\frac{M}{\mathcal{W}}\right)^{1/3}(4M^2-3Q^2)\, ,\\ 
        L_0=r_0\, \sqrt{\frac{M r_0-Q^2}{r_0^2-3 M r_0+2 Q^2}}\, &,~~
        E_0=\frac{r_0^2-2 Mr_0+Q^2}{r_0 \sqrt{r_0^2-3 M r_0+2 Q^2}}\, ,~~
        \Omega_0=\frac{\sqrt{M r_0-Q^2}}{r_0^2}\, ,
         \end{aligned}
\end{equation}
where, to avoid very long expressions, $
L_0, E_0$ and $\Omega_0$ are given implicitly in terms of $r_0$ and we introduced 
\begin{equation}
\label{W}
 \mathcal{W}=\Big(8\, M^4- 9\, M^2Q^2+2\, Q^4+\, Q^2\sqrt{5\, M^4-9\, M^2Q^2+4\, Q^4}\Big).   
\end{equation}
These are the well-known ISCO parameters of a RN black hole \cite{article,Pugliese:2010he,Pugliese:2013zma, Das:2016opi}. 
We can proceed analogously for the tidal corrections by solving \eqref{ISCO_cond} to linear order in $\langle\mathcal{E}^{\mathsf{q}}\rangle$ and we find
\begin{equation}\label{eq:corr_uncharged} 
    \begin{aligned}
    r_1&= \frac{2 r_0^2 (r_0^2-2 Mr_0+Q^2)}{M^2 C} \Big(2 M^3 \left(44 Q^2 r_0^2+63 r_0^4\right)-M^2 \left(52 Q^4 r_0+199 Q^2 r_0^3+101 r_0^5\right)\\
    &-48 M^4 r_0^3+M \left(10 Q^6+99 Q^4 r_0^2+111 Q^2 r_0^4+34 r_0^6\right)-4 \left(4 Q^6 r_0+7 Q^4 r_0^3+5 Q^2 r_0^5+r_0^7\right)\Big) \, , \\
        L_1&=\frac{1}{2 M^2 \sqrt{ M r_0-Q^2} (r_0^2 -3 Mr_0 + 2Q^2)^{3/2}}\Big(6 M^4 r_0^2 (2 r_0-r_1)\\
        &+M^3 r_0 \left(Q^2 (9 r_1-20 r_0)+r_0^2 (4 r_0+r_1)\right)+2 M^2 \left(Q^4 (4 r_0-2 r_1)-5 Q^2 r_0^3-9 r_0^5\right)\\
        &+2 M \left(6 Q^4 r_0^2+9 Q^2 r_0^4+5 r_0^6\right)-2 r_0 \left(Q^2+r_0^2\right) \left(2 Q^4+Q^2 r_0^2+r_0^4\right)\Big)\, ,
        \\
        E_1&= -\frac{1}{2 M^2 r_0^2 \left(r_0^2-3 Mr_0+2 Q^2\right)^{3/2}}\Big(M^2 r_1 \left(M r_0^2 (6 M-r_0)+4 Q^4-9 M Q^2 r_0\right)\\
        &+2 r_0 \left(r_0^2-2 Mr_0+Q^2\right) \left(r_0^2 \left(8 M^2-7 M r_0+2 r_0^2\right)+Q^2 r_0 (3 r_0-8 M)+2 Q^4\right)\Big)\, , \\
        \Omega_1&=\frac{4 M^2 Q^2 (r_1-r_0)+M^3 r_0 (4 r_0-3 r_1)-6 M Q^2 r_0^2+4 Q^2 r_0^3+4 Q^4 r_0-2 r_0^5}{2 M^2 r_0^3 \sqrt{M r_0-Q^2}}\, ,
    \end{aligned}
\end{equation}
where, for the ease of notation, all the expressions are written in terms of $r_0$ given in Eq.~\eqref{eq:unp_uncharged} and some expressions are implicitly given in terms of $r_1$. Moreover we defined
\begin{eqnarray}
\label{den}
    C&=&Q^4 r_0^2 \left(179 M r_0-222 M^2-24 r_0^2\right)+6 Q^6 r_0 (17 M-8 r_0)-16 Q^8\cr
    &+&12 M Q^2 r_0^3 \left(17 M^2-18 M r_0+4 r_0^2\right)+M r_0^4 \left(96 M^2 r_0-72 M^3-34 M r_0^2+3 r_0^3\right).
\end{eqnarray}
The tidal corrections to the ISCO parameters are plotted
in Fig.~\ref{unc_corrections} as a function of the black hole's charge-to-mass ratio.
These results put in solid grounds what could have been guessed from intuition based on the background solution. As the charge increases, the black hole's throat elongates dragging with it the ISCO, which gets closer to the hole. Thus, one expects that the tidal effects on the ISCO due to an external source should be washed away. This is precisely what can be seen in Fig.~\ref{unc_corrections}, where the magnitude of all tidal corrections decreases as a function of the hole's charge. A particularly interesting case is the extremal limit ($Q/M=1$). In that situation the hole's throat is infinitely long, but we find that tidal corrections persist and converge to 
\begin{equation}
   r_1= 648 M , \quad L_1=-126 \sqrt{2} M , \quad  E_1=-\sqrt{\frac{3}{2}}\frac{105}{4} , \quad \Omega_1=-\frac{73 \sqrt{3}}{4 M}\,.
\end{equation}
That is, we have found that tidal effects are suppressed as the black hole approaches extremality, but they survive the extremal limit and yield a finite correction.\footnote{The ISCO parameters for an extremal RN black hole, in the absence of tidal deformation simplify to $ r_0=4\,M,\quad E_0=\sqrt{\frac{3}{2}}\frac{3}{4},\quad L_0= 2\sqrt{2}\,M,\quad \Omega_0=\frac{\sqrt{3}}{16\,M}.$} 

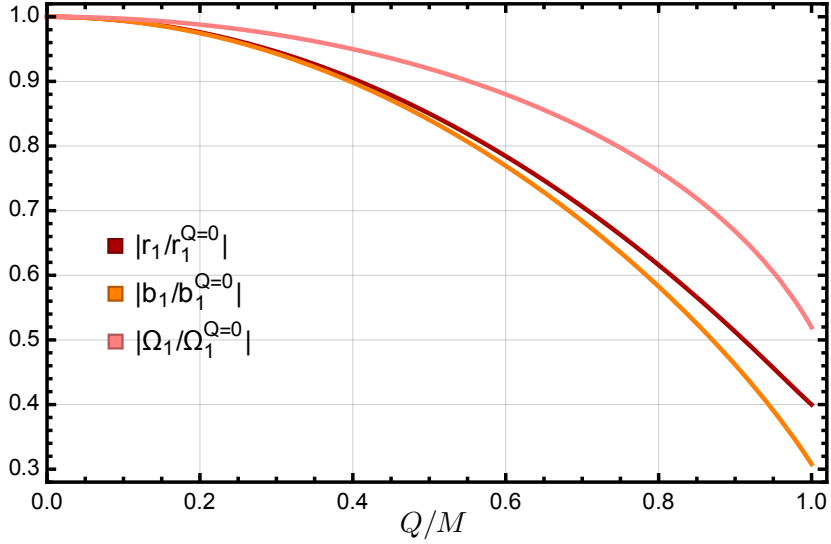
\begin{figure*}
    \centering
    \includegraphics[width=0.70\textwidth]{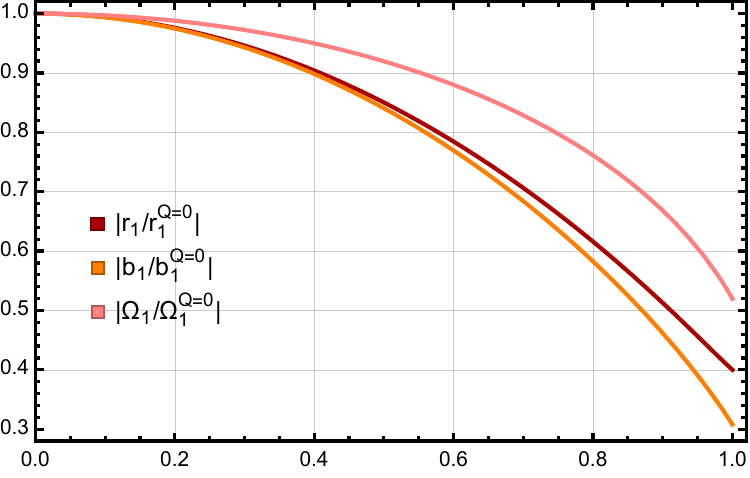}
    \begin{picture}(0,0)
        \put(-165,-4){$Q/M$ }
    \end{picture}
    \caption{Tidal corrections to the light ring's position, impact parameter and orbital frequency as a function of the black hole's charge-to-mass ratio $Q/M$. The corrections are normalised to their values in the neutral black hole case.}
    \label{lr_corrections}
\end{figure*}

One finds similar conclusions by examining the corrections to the light ring, characterised by its position $r_{\rm LR}$ and impact parameter $b_{\rm LR}=L/E$. Using the averaged Hamiltonian \eqref{averageH}, these are found by solving
\begin{equation}
    \label{eq:lr_cond}
    \langle H\rangle \vert_{r=r_{\rm LR}}=0~,
    ~~
    \frac{d\langle H\rangle}{dr}\bigg\vert_{r=r_{\rm LR}}=0~.
\end{equation}
Proceeding perturbatively as before, we search for solutions of the form 
\begin{equation}
    \label{eq:lr_exp}
    r_{\rm LR}\simeq r_0-\frac{M^{2}}{2} \langle\mathcal{E}^{\mathsf{q}}\rangle\, r_1\, , \quad b_{\rm LR}\simeq b_0-\frac{M^{2}}{2} \langle\mathcal{E}^{\mathsf{q}}\rangle\, b_1\, , \quad \Omega_{\rm LR}\simeq \Omega_0-\frac{M^{2}}{2} \langle\mathcal{E}^{\mathsf{q}}\rangle \, \Omega_1\,.
\end{equation}
The solutions at zeroth order in $\langle\mathcal{E}^{\mathsf{q}}\rangle$ are of course those of a RN black hole \cite{Tsukamoto:2021fsz}
\begin{equation}
    \label{eq:LR_unp}
    r_0=\frac{3 M+\sqrt{9 M^2-8 Q^2}}{2} ,~~ b_0=\frac{(3 M+\sqrt{9 M^2-8 Q^2})^2}{2[2M(3 M+\sqrt{9 M^2-8 Q^2})-4Q^2]^{1/2}},~~ \Omega_0=\frac{1}{b_0},
\end{equation}
while at first order in $\langle\mathcal{E}^{\mathsf{q}}\rangle$ we obtain
\begin{equation}
\begin{aligned}  \label{eq:lr_corrections}
        r_1&=-15M+9\frac{Q^2}{M}+\frac{Q^2}{M^2}\sqrt{9M^2-8Q^2}-\frac{45M^2-38Q^2}{\sqrt{9M^2-8Q^2}}\, ,
        \\  \\
        b_1&=\frac{\sqrt{2M(3 M+\sqrt{9 M^2-8 Q^2})-4Q^2}}{2M^2}(5M(3 M+\sqrt{9 M^2-8 Q^2})-4Q^2)\, ,   \\  \\
        \Omega_1&=\frac{4 (Q^2-M (\sqrt{9 M^2-8 Q^2}+2 M)) \sqrt{2 M (\sqrt{9 M^2-8 Q^2}+3 M)-4 Q^2}}{M^2 (\sqrt{9 M^2-8 Q^2}+3 M)^2}\, . 
\end{aligned}
\end{equation}
Note that both the tidal corrections presented in Eq.~\eqref{eq:lr_corrections} and the unperturbed values \eqref{eq:LR_unp} for the light ring, reduce to the ones obtained for a Schwarzschild black hole in Ref.~\cite{Cardoso_2021} in the limit $Q \rightarrow 0$, as expected. In Fig.~\ref{lr_corrections} we plot the behaviour of the tidal corrections as a function of $Q/M$. Similarly to the ISCO corrections, we see that tidal effects are monotonically decreasing functions of $Q/M$ that converge to a non-zero value at extremality. In the case of extremal black holes the light ring parameters simplify to 
\begin{equation}
    \label{eq:extremalISCO_LR}
    \begin{split}
    r_{0}&=2M\, , \qquad b_0=4M\, , \qquad \Omega_0=\frac{1}{4M}\, , \\
    r_1&=- 12 M\, , \quad b_1=16 M \, , \quad \Omega_{1}=-\frac{1}{M}.
    \end{split}
\end{equation}

\subsection{The Case of a Charged Particle}

Here we consider the possibility that the small object also carries electric charge $q$. In that case, besides probing the spacetime geometry, the particle also interacts with the electromagnetic field of the tidally deformed background. Although in our set up there is no electromagnetic tidal source, the Maxwell field strength is coupled via the hole's charge to the gravitational deformation. Therefore, in this case the tidal correction to the charged particle's trajectory contains a gravitational and an electromagnetic contribution. Our next goal is to quantify these effects.

Focusing again on quasi circular orbits, the averaged Hamiltonian for a point charge is
\begin{equation}\label{H_carica}
    \begin{aligned}
\left \langle H \right \rangle &= -\frac{1}{2} \Biggl( \frac{\left(  \frac{\tilde{q}Q}{r}-\langle E\rangle\right)^2}{f}-\frac{\left \langle L \right \rangle^2}{r^2} \Biggr)+\frac{r^{2}}{2}\langle\mathcal{E}^{\mathsf{q}}\rangle\,  \Biggl[ \left(  \frac{\tilde{q}Q}{r}- \langle E\rangle  \right)^2\, + f\frac{\left \langle L \right \rangle^2}{r^2} \Biggr]\\ 
&\sim -\frac{1}{2} \Biggl( \frac{\langle E\rangle  \bigl( \langle E \rangle -\frac{2 \tilde{q}Q}{r}\bigr)}{ f}-\frac{\left \langle L \right \rangle^2}{r^2} \Biggr)+\frac{r^{2}}{2}\langle\mathcal{E}^{\mathsf{q}}\rangle\,  \Biggl[  \langle E\rangle \left(\langle E\rangle-\frac{2\tilde{q}Q}{r}\right) \, + f\frac{\left \langle L \right \rangle^2}{r^2} \Biggr]\, ,
    \end{aligned}
\end{equation}
where $\tilde{q}=q/m$ is the particle's charge-to-mass ratio, and the last approximation holds for $\tilde{q}\ll1$. This regime excludes elementary particles, whose $\tilde{q}$ in geometric units is very large ($\sim10^{21}$), but is otherwise valid for macroscopic bodies. Here we should restrict to that limit for simplicity, and work at first order in $\tilde{q}$. The ISCO parameters are written as

\begin{equation}
    \label{tidal_charge}
    \begin{aligned}
       r_{\rm ISCO}&\simeq r_0+\tilde{q}r_0^q-\frac{M^2}{2}\left \langle \mathcal{E}^{\mathsf{q}} \right \rangle (r_1+\tilde{q}r_1^q)\,, \quad L_{\rm ISCO}\simeq L_0+\tilde{q}L_0^q-\frac{M^2}{2}\left \langle \mathcal{E}^{\mathsf{q}} \right \rangle(L_1+\tilde{q}L_1^q)\,,  \\
       \quad  E_{\rm ISCO}&\simeq E_0+\tilde{q}E_0^q-\frac{M^2}{2}\left \langle \mathcal{E}^{\mathsf{q}} \right \rangle(E_1+\tilde{q}E_1^q)\,,\quad \Omega_{\rm ISCO}\simeq \Omega_0+\tilde{q}\Omega_0^q-\frac{M^2}{2}\left \langle \mathcal{E}^{\mathsf{q}} \right \rangle(\Omega_1+\tilde{q}\Omega_1^q)\, ,
    \end{aligned}
\end{equation}
where $(r_{0,1}, L_{0,1}, E_{0,1},\Omega_{0,1})$ are the gravitational contribution given in Eqs.~\eqref{eq:unp_uncharged} and \eqref{eq:corr_uncharged}, while the superscript $q$ refers to the electromagnetic contribution and each piece can be obtained proceeding as in the sections above. Their analytical expressions read
 \begin{equation}
 \label{qcorr}
     \begin{split}
         r_0^q&=-\frac{Qr_0\sqrt{r_0^2-3 M r_0+2 Q^2} \left( r_0^2-6 M r_0+6 Q^2\right) \left(r_0^2-2 M r_0+Q^2\right)^2}{C},
         \\
         L_0^q&= \frac{2Q r_0 \left( r_0^2-2 M r_0+Q^2\right)}{C \sqrt{M r_0-Q^2}} \Big(5 Q^6+M Q^2 r_0^2 (27 M-14 r_0)+7 Q^4 r_0 (r_0-3 M)
         \\
         &~~~~~~-M r_0^3 \left(12 M^2-9 M r_0+r_0^2\right)\Big),
         \\
         E_0^q&=\frac{Q}{r_0 C}\Big(2 M Q^2 r_0^3 \left(36 M^2-44 M r_0+9 r_0^2\right)-Q^4 r_0^2 \left(84 M^2-81 M r_0+10 r_0^2\right)
         \\
         &~~~~~~+M r_0^4 \left(36 M^2 r_0-24 M^3-12 M r_0^2+r_0^3\right)+8 Q^6 r_0 (5 M-3 r_0)-6 Q^8\Big),
         \\
         \Omega_0^q&=\frac{2 Q\left(r_0^2-3 M r_0+2 Q^2\right)^{3/2} \left(M^2 r_0^4-Q^2 r_0^2 \left(3 M^2-M r_0+r_0^2\right)+3 M Q^4 r_0-Q^6\right)}{Cr_0^2 \sqrt{M r_0-Q^2} },
     \end{split}
 \end{equation}
where, to avoid heavy notation, we chose to give the various expressions in terms of $r_0$ which is given in Eq.~\eqref{eq:unp_uncharged}. For the tidal corrections we find
\begin{equation}
\label{qcorrtidal}
\begin{split}
 r_1^q&=\frac{r_0}{C^2  M^2\sqrt{\mathcal{K}}} \left\{\xi _1 \left[M^2 \left(199 Q^2 r_0^3+52 Q^4 r_0+101 r_0^5\right)-2 M^3 \left(44 Q^2 r_0^2+63 r_0^4\right)\right. \right.
 \\
 &\left.\left.+48 M^4 r_0^3-M \left(99 Q^4 r_0^2+111 Q^2 r_0^4+10 Q^6+34 r_0^6\right)+4 \left(5 Q^2 r_0^5+7 Q^4 r_0^3+4 Q^6 r_0+r_0^7\right)\right]\right.
 \\
 &\left.-C Q \left(r_0^2-2 Mr_0+Q^2\right)\xi _2\right\},\\
 L_1^q&=\frac{Q}{4 C \mathcal{K}^2 M^2 r_0 \left(M r_0-Q^2\right)^{3/2} \left(r_0^2-2 Mr_0+Q^2\right)}\left(\xi _3+\frac{2}{\sqrt{\mathcal{K}}}\xi _4\right),\\
 E_1^q&=\frac{Q\left(\xi _5+\sqrt{\mathcal{K}} \xi _6\right)}{4 r_0^3 \left(r_0^2-2 Mr_0+Q^2\right)},\\
 \Omega_1^q&=\frac{\xi _7}{C M^2  r_0^4 \sqrt{M r_0-Q^2} \left(r_0^2-2 Mr_0+Q^2\right)^2},
\end{split}
\end{equation}
 where $\mathcal{K}= r_0^2-3 M r_0+2 Q^2$ and the explicit expression for $\xi_{1,..,7}$ is relegated to App.~\ref{app:xi}. Moreover, $C$ is defined in Eq.~\eqref{den}.
\begin{figure*}
    \centering
    \includegraphics[width=0.70\textwidth]{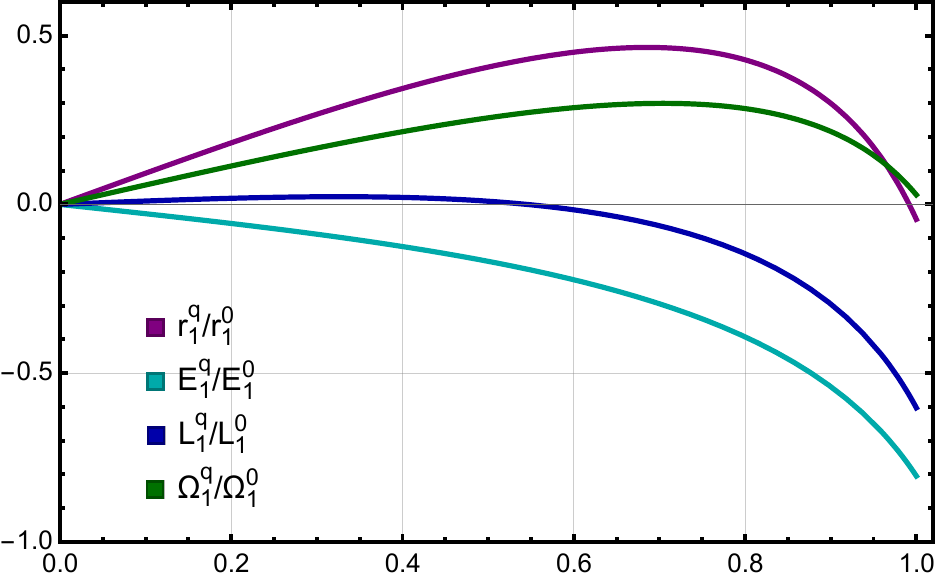}
    \begin{picture}(0,0)
        \put(-160,-5){$Q/M$ }
    \end{picture}
    \caption{Ratio between the electromagnetic and gravitational contributions to the tidal corrections of the ISCO parameters, as a function of $Q/M$. }
    \label{ratio}
\end{figure*}
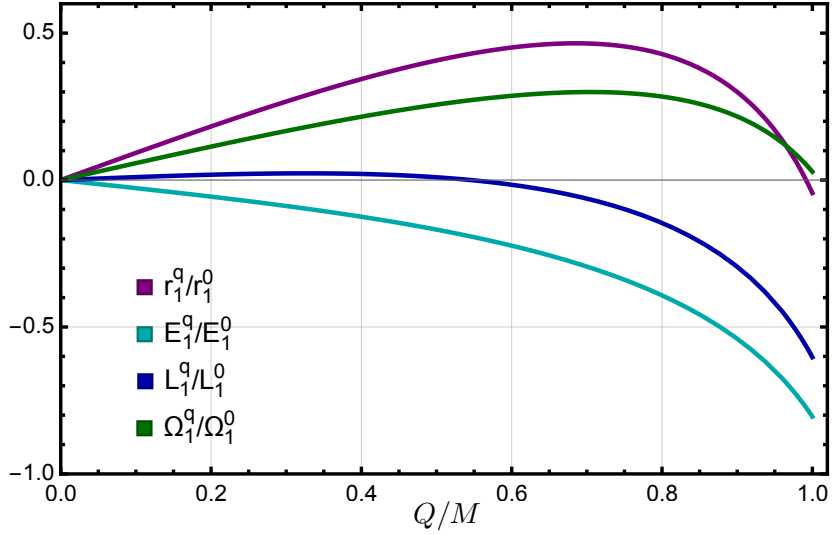

In the extremal limit ($Q=M$), the ISCO parameters up to first order in $\tilde q$ simplify to 
\begin{equation}
    \label{eq:ISCO_RN_charge}
    \begin{split}
    r_{\rm ISCO}^{\rm extremal}&\simeq 4 M-\sqrt{\frac{2}{3}} \tilde{q}M -\frac{M^3}{2}\left \langle \mathcal{E}^{\mathsf{q}} \right \rangle \Big(648-\tilde{q}\, 35 \sqrt{\frac{2}{3}}\Big), \\ 
    L_{\rm ISCO}^{\rm extremal}& \simeq 2 \sqrt{2} M-\sqrt{3} \tilde{q}M +\frac{M^3}{2}\left \langle \mathcal{E}^{\mathsf{q}} \right \rangle \Big(126\sqrt{2}-\tilde{q}\, 62\sqrt{3}\Big), \\ E_{\rm ISCO}^{\rm extremal}& \simeq \sqrt{\frac{3}{2}}\frac{3 }{4}+\frac{\tilde{q}}{16 }+\frac{M^2}{2}\left \langle \mathcal{E}^{\mathsf{q}} \right \rangle \Big(\frac{105}{4}\sqrt{\frac{3}{2}}-\tilde{q}\,\frac{207}{8}\Big)\,  , \\ \Omega_{\rm ISCO}^{\rm extremal}& \simeq\, \frac{\sqrt{3}}{16\,M}\,-\tilde{q}\frac{\sqrt{2}}{96\, M}+\frac{M}{2}\left \langle \mathcal{E}^{\mathsf{q}} \right \rangle \Big(\frac{73}{4}\sqrt{3}+\tilde{q}\frac{61}{96}\sqrt{2}\Big).
    \end{split}
\end{equation}

Denoting with $\mathcal{A}^{0}_{1}$ and $\mathcal{A}^{q}_{1}$ respectively the gravitational and electromagnetic contributions to the tidal deformations [with $\mathcal{A}=(r,L,E,\Omega)]$, we can plot their ratios $\mathcal{A}^{q}_{1}/\mathcal{A}^{0}_{1}$ as in Fig.~\ref{ratio}. In this way, it is possible to assess the relative importance of each effect, as a function of the black hole's charge-to-mass ratio.

 The electromagnetic contribution of course vanishes for $Q=0$, and it is activated as $Q$ increases. While the behaviour is slightly different for each quantity, in all cases the gravitational correction dominates in magnitude over the electromagnetic one for all $Q$. This may have been anticipated given that we are only considering a gravitational tidal source. It is also worth noticing that, for some values of $Q$, the electromagnetic corrections to the ISCO's position and angular momentum vanish. Such a non-monotonic behaviour reveals a nontrivial balance of forces governing the particle's dynamics, which is much richer than in the neutral case.

\section{Discussion}
\label{Sec:Discussion}

In this paper we have considered the dynamics of a charged black hole EMRI subject to an external gravitational tide. This could be thought of as a hierarchical triple system, where the charged EMRI is under the influence of a much larger, neutral black hole. To that aim, we first constructed the most general stationary gravitational tidal field on a black hole carrying both electric and magnetic charges. This is represented by a time independent solution to the Einstein--Maxwell equations, linearised about the dyonic RN geometry, which asymptotically approaches a purely gravitational tide (in particular, this implies that we are not considering electromagnetic tidal sources). Next, we considered the motion of a test particle in the vicinity of the tidally deformed black hole. Focusing on quasi-circular trajectories, we constructed the averaged Hamiltonian which describes the secular modifications to the orbit due to the tidal deformation. This allowed us to obtain the tidal corrections to the ISCO and the light ring in terms of explicit functions of the binary's parameters (the masses and charges of the two bodies). 

Our results indicate that tidal effects are suppressed as the black hole charge $Q$ increases. This can be interpreted physically from the fact that, as $Q$ approaches its extremality bound $M$, the black hole's throat elongates and becomes infinite precisely at extremality $Q=M$. This elongation drags both the ISCO and the light ring towards smaller radii and, consequently, the tidal effects due to external sources decrease. However, they survive in the extremal limit and converge to a finite, non-zero result. If, in addition, the particle carries electric charge then tidal effects also contain an electromagnetic contribution. While this incorporates a more intricate interplay of forces, the gravitational contribution dominates in the cases considered here, where the tidal source is purely gravitational. 

Some of our conclusions may still hold in the case that the charged black hole is replaced by a neutral, rotating one. From our results, one would expect that tidal effects on the dynamics of nearby matter are suppressed as the black hole spin increases (at least for matter close to the ISCO and waves propagating close to the light ring). A natural continuation of our work is precisely addressing that computation. Similarly, the solution constructed in Sec.~\ref{Sec:Tides} and the methods to study binary mechanics of Secs. \ref{Dynamics} and \ref{ISCOstuff} pave the way towards exploring tidal effects on other charged systems. A relevant case would be that of topological stars \cite{Bah:2020ogh}, since these are well-defined candidates of horizonless compact objects. As such, their response to tidal sources is expected to differ significantly from that of black holes, and elucidating how this translates into the orbital dynamics of nearby matter is an interesting future direction. Another intriguing direction is departing from the quasi circular approximation. For example, in \cite{Camilloni:2023xvf} some of us studied recently how the eccentricity of a black hole binary system is influenced by the tidal field of a third body, accounting for strong field effects (beyond the pure PN framework). There, it was shown that the external tidal field can considerably increase the eccentricity of the binary system and accelerate the merger, compared to the standard Newtonian description. It would be very interesting to understand how this changes in the presence of an additional, non-gravitational radiation channel, such as the electromagnetic one, by allowing the black holes in the binary system to carry electric charge. We intend to address these issues in future research.

\section*{Acknowledgements}

We thank V. Cardoso, T. Harmark and T. Katagiri for their interesting comments and suggestions on the manuscript. Elisa Grilli, Marta Orselli, and Daniele Pica acknowledge financial support from the Italian Ministry of University and Research (MUR) through the program “Dipartimenti di Eccellenza 2018-2022” (Grant SUPER-C). Elisa Grilli, Marta Orselli, and Daniele Pica also acknowledge support from ``Fondo di Ricerca d'Ateneo" 2021 (MEGA) and 2023 (GraMB) of the University of Perugia. Marta Orselli acknowledges support from the Italian Ministry of University and Research (MUR) via the PRIN 2022ZHYFA2, GRavitational wavEform models for coalescing compAct binaries with eccenTricity (GREAT). David Pereñiguez acknowledges financial support by the VILLUM Foundation (grant no. VIL37766) and the DNRF Chair program (grant no. DNRF162) by the Danish National Research Foundation.

\appendix 

\section{Explicit Solutions for $\mathcal{X}_{\ell,m},\mathcal{U}_{\ell,m},\mathcal{V}_{\ell,m},\Phi^{\pm}_{\ell,m}$}\label{AppExpSol}

Setting the scale to $M=1$ and, without loss of generality, choosing the normalisations $\mathcal{X}=r^{\ell}\left(1+O(r^{\ell-1})\right)$ and $\mathcal{V}=r^{\ell-1}\left(1+O(r^{\ell-2})\right)$, the solutions for $\ell=2,3,4$ read:
\begin{itemize}
    \item $\ell=2$:
        \begin{align}
        \mathcal{X}_{2,m}(r)&= \frac{\left(\mathcal{Q}^2+r^2-2 r\right)^2}{r^2} \, \\
        \mathcal{U}_{2,m}(r)&= \mathcal{Q}^2+r^2-2\, \\ 
        \mathcal{V}_{2,m}(r)&=\frac{\mathcal{Q}^2}{r}+r-2\, \\
        \Phi^{+}_{2,m}(r)&= \frac{1}{2} \left(\mathcal{Q}^2+3 r^2-4\right) \, \\
        \Phi^{-}_{2,m}(r)&= -\frac{3}{2}  \left(r^2-\mathcal{Q}^2\right)\,
    \end{align}
   
    \item $\ell=3$:    
    \begin{align}
        \mathcal{X}_{3,m}(r)&= \frac{\left(\mathcal{Q}^2+r^2-2 r\right)^2 \left(\mathcal{Q}^2+5 r^2-5 r\right)}{5 r^3}\, \\
        \mathcal{U}_{3,m}(r)&=\frac{1}{5} \left(-6 \mathcal{Q}^2+5 r^3-10 r^2+\frac{\mathcal{Q}^4}{r}+6 \mathcal{Q}^2 r+4\right) \, \\
        \mathcal{V}_{3,m}(r)&=\frac{-\mathcal{Q}^6+15 r^6-50 r^5+\left(21 \mathcal{Q}^2+40\right) r^4-32 \mathcal{Q}^2 r^3+5 \mathcal{Q}^4 r^2+2 \mathcal{Q}^4 r}{15 r^4}\, \\
        \Phi^{+}_{3,m}(r)&=\frac{2}{5} \left(5 r^3-9 r^2+\mathcal{Q}^2 (3 r-1)+2\right) \, \\
        \Phi^{-}_{3,m}(r)&=-\frac{4 \left(-\mathcal{Q}^4+5 r^4-8 r^3+4 \mathcal{Q}^2 r\right)}{5 r} \,
    \end{align}
  
    \item $\ell=4$:
    \begin{align}
        \mathcal{X}_{4,m}(r)&= \frac{\left(\mathcal{Q}^2+r^2-2 r\right)^2 \left(-2 \mathcal{Q}^2+7 r^3-14 r^2+3 \left(\mathcal{Q}^2+2\right) r\right)}{7 r^3}\, \\
        \mathcal{U}_{4,m}(r)&= \frac{-6 \mathcal{Q}^4+21 r^5-70 r^4+30 \left(\mathcal{Q}^2+2\right) r^3-60 \mathcal{Q}^2 r^2+\left(9 \mathcal{Q}^4+24 \mathcal{Q}^2-8\right) r}{21 r} \\\notag
        \mathcal{V}_{4,m}(r)&=\frac{\mathcal{Q}^6}{14 r^4}-\frac{\mathcal{Q}^4}{7 r^3}+r^3-\frac{9 \mathcal{Q}^4}{14 r^2}-\frac{9 r^2}{2}+\frac{5 \left(\mathcal{Q}^2+4\right) \mathcal{Q}^2}{7 r}+\frac{3}{7} \left(4 \mathcal{Q}^2+15\right) r \\
        &-\frac{5}{14} \left(13 \mathcal{Q}^2+8\right)\, \displaybreak \\
        \Phi^{+}_{4,m}(r)&=\frac{1}{42}  \left(9 \mathcal{Q}^4+12 \mathcal{Q}^2+105 r^4-320 r^3+30 \left(3 \mathcal{Q}^2+8\right) r^2-120 \mathcal{Q}^2 r-16\right)  \, \\
        \Phi^{-}_{4,m}(r)&=-\frac{5 \left(4 \mathcal{Q}^4+21 r^5-60 r^4+10 \left(\mathcal{Q}^2+4\right) r^3-3 \mathcal{Q}^2 \left(\mathcal{Q}^2+4\right) r\right)}{14 r} \,
    \end{align}
\end{itemize}
Similar solutions are found for $\ell>4$.

\section{Explicit Expressions for $\xi_{1,...,7}$}
\label{app:xi}
Here we write explicitly the expressions for $\xi_1,...,\xi_7$ introduced in Eqs.~\eqref{qcorrtidal}. Their expressions are given in terms of $r_0$, $r_0^q$, $r_1$, $L_1$ and $E_1$ (see Eqs.~\eqref{eq:unp_uncharged}~\eqref{eq:corr_uncharged} and \eqref{qcorr}). In detail we have 
    \begin{align}
    \notag
    \xi_1&=\frac{4 Q r_0}{C} \left\{2 \mathcal{K} \left(5 Q^6-M r_0^3 \left(12 M^2-9 M r_0+r_0^2\right)+M Q^2 r_0^2 (27 M-14 r_0)+7 Q^4 (r_0^2-3 Mr_0)\right)\right.
        \\ \notag
        &\left.\times \left(r_0^2 \left(24 M^2-16 M r_0+3 r_0^2\right)+3 Q^2 r_0 (3 r_0-10 M)+10 Q^4\right) \left(r_0^2-2 Mr_0+Q^2\right)^3\right.
        \\ \notag
        &\left.+\mathcal{K} \left[Q^4 r_0^2 \left(84 M^2-81 M r_0+10 r_0^2\right)-2 M Q^2 r_0^3 \left(36 M^2-44 M r_0+9 r_0^2\right)\right.\right.
        \\ \notag
        &\left.\left.+M r_0^4 \left(24 M^3-36 M^2 r_0+12 M r_0^2-r_0^3\right)+8 Q^6 r_0 (3 r_0-5 M)+6 Q^8\right]\right.
        \\ \notag
        &\left.\times \left[4 Q^4 r_0^2 \left(22 M^2-24 M r_0+3 r_0^2\right)-Q^2 r_0^3 \left(72 M^3-96 M^2 r_0+14 M r_0^2+3 r_0^3\right)\right.\right.
        \\ \notag
        &\left.\left.+4 M^2 r_0^4 \left(6 M^2-10 M r_0+3 r_0^2\right)+Q^6 r_0 (29 r_0-42 M)+6 Q^8\right]\right\}
        \\ \notag
        &-Q^7 r_0^2 \left(2 \sqrt{\mathcal{K}} \frac{r_0^q}{Q} \left(468 M^2-625 M r_0+72 r_0^2\right)+r_0 \left(54 M^2+9 M r_0+80 r_0^2\right)\right)
        \\ \notag
        &+Q^5 r_0^3 \left[2 \sqrt{\mathcal{K}} \frac{r_0^q}{Q} \left(864 M^3-1400 M^2 r_0+361 M r_0^2-12 r_0^3\right)\right.
        \\ \notag
        &\left.+r_0 \left(420 M^3-676 M^2 r_0+645 M r_0^2-148 r_0^3\right)\right]
        \\ \notag
       &+Q^3 r_0^4 \left[2 \sqrt{\mathcal{K}} M \frac{r_0^q}{Q} \left(1524 M^2 r_0-792 M^3-668 M r_0^2+87 r_0^3\right)\right.
        \\ \notag
        &\left.+r_0 \left(1068 M^3 r_0-1094 M^2 r_0^2-504 M^4+413 M r_0^3-48 r_0^4\right)\right]
        \\ \notag
        &+M Q r_0^5 \left[2 \sqrt{\mathcal{K}} \frac{r_0^q}{Q} \left(412 M^2 r_0^2-648 M^3 r_0+288 M^4-104 M r_0^3+9 r_0^4\right)\right.
        \\ \notag
        &\left.+r_0 \left(376 M^2 r_0^2-336 M^3 r_0+144 M^4-168 M r_0^3+25 r_0^4\right)\right]
        \\ 
        &+2 Q^9 r_0 \left(2 \sqrt{\mathcal{K}} \frac{r_0^q}{Q} (59 M-58 r_0)+3 r_0 (10 r_0-7 M)\right)+8 Q^{11} \left(r_0-2 \sqrt{\mathcal{K}} \frac{r_0^q}{Q}\right),
        \\ \notag
        \xi_2&=\frac{8 \mathcal{K} r_0 \left(r_0^2-2 Mr_0+Q^2\right)}{C} \left\{24 M^5 \left(23 Q^2 r_0^4+21 r_0^6\right)-6 M^4 \left(295 Q^2 r_0^5+144 Q^4 r_0^3+85 r_0^7\right)\right.
        \\ \notag
        &\left.+M^3 \left(1512 Q^2 r_0^6+2531 Q^4 r_0^4+660 Q^6 r_0^2+201 r_0^8\right)-144 M^6 r_0^5\right.
        \\ \notag
        &\left.-2 M^2 \left(219 Q^2 r_0^7+900 Q^4 r_0^5+870 Q^6 r_0^3+118 Q^8 r_0+17 r_0^9\right)\right.
        \\ \notag
        &\left.+M \left(543 Q^8 r_0^2+956 Q^6 r_0^4+371 Q^4 r_0^6+42 Q^2 r_0^8+30 Q^{10}+2 r_0^{10}\right)\right.
        \\ \notag
        &\left.-4 Q^4 r_0 \left(46 Q^4 r_0^2+27 Q^2 r_0^4+14 Q^6+5 r_0^6\right)\right\}
        \\ \notag
        &+2 \sqrt{\mathcal{K}} \frac{r_0^q}{Q} \left[-8 M^3 \left(61 Q^2 r_0^2+102 r_0^4\right)+M^2 \left(1187 Q^2 r_0^3+272 Q^4 r_0+709 r_0^5\right)+288 M^4 r_0^3\right.
        \\ \notag \displaybreak
        &\left.-M \left(549 Q^4 r_0^2+723 Q^2 r_0^4+50 Q^6+254 r_0^6\right)+8 \left(17 Q^2 r_0^5+22 Q^4 r_0^3+10 Q^6 r_0+4 r_0^7\right)\right]
        \\ \notag
        &+\left(r_0^2-2 Mr_0+Q^2\right) \left[Q^2 r_0^2 \left(158 M r_0-60 M^2-45 r_0^2\right)\right.
        \\
        &\left.+r_0^3 \left(48 M^3-122 M^2 r_0+61 M r_0^2-8 r_0^3\right)-44 Q^4 r_0^2+12 Q^6\right], \\ \notag
        \xi_3&=C r_0 \left(Q^2-M r_0\right) \left\{M^2 r_1 \left[r_0^3 \left(12 M^3-20 M^2 r_0+17 M r_0^2-4 r_0^3\right)\right.\right.
        \\ \notag
        &\left.\left.+6 M Q^2 r_0^2 (2 M-3 r_0)+Q^4 r_0 (16 r_0-19 M)+4 Q^6\right]\right.
        \\ 
        &\left.-2 r_0 \left(r_0^2-2 Mr_0+Q^2\right)^2 \left(r_0^2 \left(8 M^2-13 M r_0+4 r_0^2\right)+Q^2 r_0 (5 r_0-2 M)-2 Q^4\right)\right\}, \\
   \notag
        \xi_4&=\left(Q^2-M r_0\right) \mathcal{K}\left\{C r_0 \left(r_0^2-2 M r_0+Q^2\right) \left[2 (3 r_0-2M) \left(M r_0-Q^2\right) \frac{r_0^q}{Q} r_1 M^2\right.\right.
        \\ \notag
        &\left.\left.-\left(Q^4+r_0 (3 r_0-4 M) Q^2+2 M (M-r_0) r_0^2\right) \frac{r_1^q}{Q} M^2-2 (2 M-3 r_0) r_0 \left(r_0^2-2 M r_0+Q^2\right)^2 \frac{r_0^q}{Q}\right.\right.
        \\ \notag
        &\left.\left.+8 (M-r_0)^2 r_0^2 \left(r_0^2-2 M r_0+Q^2\right) \frac{r_0^q}{Q}\right]-C r_0 \left(Q^2-M r_0\right) \left(r_0^2-2 M r_0+Q^2\right) \right.
        \\ \notag
        &\left.\times \left[8 (r_0^q r_1+r_0 (r_1^q-3 r_0^q)) \frac{M^3}{Q}+\left((8 r_0^q-5 r_1^q) Q+3 \frac{r_0}{Q} (4 r_0 r_0^q-2 r_1 r_0^q-r_0 r_1^q)\right) M^2\right.\right.
        \\ \notag
        &\left.\left.+4 r_0 \left(3 Q^2+2 r_0^2\right) \frac{r_0^q}{Q} M-4 Q \left(Q^2+3 r_0^2\right) r_0^q\right]+2 \left[\left(Q^4+r_0 (3 r_0-4 M) Q^2+2 M (M-r_0) r_0^2\right) r_1 M^2\right.\right.
        \\ \notag
        &\left.\left.+2 (M-r_0) r_0^2 \left(r_0^2-2 M r_0+Q^2\right)^2\right] \left[2 r_0 \left(6 Q^8+8 r_0 (3 r_0-5 M) Q^6+r_0^2 \left(84 M^2-81 r_0 M+10 r_0^2\right) Q^4\right.\right.\right.
        \\ \notag
        &\left.\left.\left.-2 M r_0^3 \left(36 M^2-44 r_0 M+9 r_0^2\right) Q^2+M r_0^4 \left(24 M^3-36 r_0 M^2+12 r_0^2 M-r_0^3\right)\right) \sqrt{\mathcal{K}}\right.\right.
        \\ \notag
        &\left.\left.-3 C \left(r_0^2-2 M r_0+Q^2\right) \frac{r_0^q}{Q}\right]-2 \left[4 r_0 (3 r_0-2 r_1) M^3+\left((5 r_1-8 r_0) Q^2+r_0^2 (3 r_1-4 r_0)\right) M^2\right.\right.
        \\ \notag
        &\left.\left.-2 \left(r_0^4+3 Q^2 r_0^2\right) M+4 Q^2 r_0 \left(Q^2+r_0^2\right)\right] \left[r_0 \left(16 Q^{10}+2 r_0 (29 r_0-64 M) Q^8\right.\right.\right.
        \\ \notag
        &\left.\left.\left.+r_0^2 \left(386 M^2-313 r_0 M+48 r_0^2\right) Q^6+r_0^3 \left(631 r_0 M^2-564 M^3-184 r_0^2 M+14 r_0^3\right) Q^4\right.\right.\right.
        \\ \notag
        &\left.\left.\left.+M r_0^4 \left(408 M^3-572 r_0 M^2+240 r_0^2 M-33 r_0^3\right) Q^2\right.\right.\right.
        \\ \notag
        &\left.\left.\left.+M r_0^5 \left(204 r_0 M^3-120 M^4-116 r_0^2 M^2+27 r_0^3 M-2 r_0^4\right)\right) \sqrt{\mathcal{K}}\right.\right.
        \\ \notag
        &\left.\left.-C \left(Q^2-M r_0\right) \left(Q^2+r_0 (3 r_0-4 M)\right) \frac{r_0^q}{Q}\right]\right\} -\left\{6 r_0^2 (r_1-2 r_0) M^4\right.
        \\ \notag
        &\left.-r_0 \left((9 r_1-20 r_0) Q^2+r_0^2 (4 r_0+r_1)\right) M^3+2 \left(9 r_0^5+5 Q^2 r_0^3+2 Q^4 (r_1-2 r_0)\right) M^2\right.
        \\ \notag
        &\left.-2 \left(5 r_0^6+9 Q^2 r_0^4+6 Q^4 r_0^2\right) M+2 r_0 \left(Q^2+r_0^2\right) \left(2 Q^4+r_0^2 Q^2+r_0^4\right)\right\} \left\{2 r_0 \left[r_0 (87 M-41 r_0) Q^8-11 Q^{10}\right.\right.
        \\ \notag
        &\left.\left.+r_0^2 \left(-255 M^2+209 r_0 M-29 r_0^2\right) Q^6+r_0^3 \left(360 M^3-400 r_0 M^2+106 r_0^2 M-7 r_0^3\right) Q^4\right.\right.
        \\ \notag
        &\left.\left.+M r_0^4 \left(-252 M^3+348 r_0 M^2-135 r_0^2 M+17 r_0^3\right) Q^2\right.\right.
        \\ \notag
        &\left.\left.+M r_0^5 \left(72 M^4-120 r_0 M^3+64 r_0^2 M^2-14 r_0^3 M+r_0^4\right)\right] \mathcal{K}^{3/2}-C \left(M r_0-Q^2\right) \frac{r_0^q}{Q} \left[r_0 (2 r_0-3 M) \right.\right.
        \\ 
        &\left.\left. \times \left(r_0^2-2 M r_0+Q^2\right)+\left(3 Q^2+r_0 (7 r_0-10 M)\right) \mathcal{K}\right]\right\},  \\ \notag
        \xi_5&=4 \left\{\frac{r_0}{C M^2} \left[C+Q^4 r_0^2 \left(84 M^2-81 M r_0+10 r_0^2\right)-2 M Q^2 r_0^3 \left(36 M^2-44 M r_0+9 r_0^2\right)\right.\right.
        \\ \notag
        &\left.\left.+M r_0^4 \left(24 M^3-36 M^2 r_0+12 M r_0^2-r_0^3\right)+8 Q^6 r_0 (3 r_0-5 M)+6 Q^8\right]\left[L_1 M^2 \sqrt{\mathcal{K} \left(M r_0-Q^2\right)}\right.\right.
        \\ \notag
        &\left.\left.+Q^2 \left(M^2 (r_1-2 r_0)-M r_0^2+r_0^3\right)-M r_0 \left(M^2 (r_1-2 r_0)+r_0^3\right)+Q^4 r_0\right]+\frac{r_0 r_1}{C} \left[C r_0 (M-r_0)\right.\right. 
        \\ \notag
        &\left.\left.+Q^6 r_0^2 \left(124 M^2-105 M r_0+10 r_0^2\right)+M Q^4 r_0^3 \left(169 M r_0-156 M^2-28 r_0^2\right)\right.\right.
        \\ \notag
        &\left.\left.+M Q^2 r_0^4 \left(96 M^3-124 M^2 r_0+30 M r_0^2-r_0^3\right)+M^2 r_0^5 \left(36 M^2 r_0-24 M^3-12 M r_0^2+r_0^3\right)\right.\right.
        \\ \notag \displaybreak
        &\left.\left.+2 Q^8 r_0 (12 r_0-23 M)+6 Q^{10}\right]+\frac{r_0^2}{C M^2} \left(r_0^2-2 Mr_0+Q^2\right)^2 \left[C+Q^4 r_0^2 \left(84 M^2-81 M r_0+10 r_0^2\right)\right.\right.
        \\ \notag
        &\left.\left.-2 M Q^2 r_0^3 \left(36 M^2-44 M r_0+9 r_0^2\right)+M r_0^4 \left(-36 M^2 r_0+24 M^3+12 M r_0^2-r_0^3\right)\right.\right.
        \\ \notag
        &\left.\left.+8 Q^6 r_0 (3 r_0-5 M)+6 Q^8\right]+\frac{1}{\sqrt{\mathcal{K}} M^2 Q}\left[M^2 r_0^q r_1 \left(Q^2 r_0 (3 r_0-4 M)+2 M r_0^2 (M-r_0)+Q^4\right)\right.\right.
        \\ 
        &\left.\left.-M^2 r_0 r_1^q \left(Q^2-M r_0\right) \left(r_0^2-2 Mr_0+Q^2\right)+2 r_0^2 r_0^q (M-r_0) \left(r_0^2-2 M r_0+Q^2\right)^2\right]\right\}\,, \\ \notag
        \xi_6&=\frac{4}{M^2} \left\{\frac{\left(M r_0-Q^2\right)}{\mathcal{K} Q} \left[r_0^2 M^2(4 r_0 r_0^q+r_0 r_1^q-3 r_0^q r_1)-M^2 Q^2 (8 r_0 r_0^q+r_0 r_1^q-5 r_0^q r_1)\right.\right.
        \\ \notag
        &\left.\left.+2 M^3 r_0 (r_0 (6 r_0^q+r_1^q)-4 r_0^q r_1)-2 M r_0^2 r_0^q \left(3 Q^2+r_0^2\right)+4 Q^2 r_0 r_0^q \left(Q^2+r_0^2\right)\right]\right.
        \\ \notag
        &\left.-\frac{2 r_0\left(r_0^2-2 Mr_0+Q^2\right)^2}{C \sqrt{ (M r_0- Q^2)\mathcal{K}}}  \left[M Q^2 r_0^2 (27 M-14 r_0)-M r_0^3 \left(12 M^2-9 M r_0+r_0^2\right)\right.\right.
        \\ \notag
        &\left.\left.+7 Q^4 r_0 (r_0-3 M)+5 Q^6\right] \times \left[2 r_0 \left(Q^2+r_0^2\right) \sqrt{M r_0-Q^2}-M^2 \left(\sqrt{\mathcal{K}} L_1+2 (2 r_0-r_1) \sqrt{M r_0-Q^2}\right)\right]\right\}
        \\ \notag
        &-\frac{8 L_1 r_0^q \sqrt{\mathcal{K} \left(M r_0-Q^2\right)}}{Q}-\frac{1}{M^2 \mathcal{K}^2}\left\{\frac{r_0}{\sqrt{\mathcal{K}}} \left[M^2 \left(4 Q^6+r_0 (16 r_0-19 M) Q^4+6 M r_0^2 (2 M-3 r_0) Q^2\right.\right.\right.
        \\ \notag
        &\left.\left.\left.+r_0^3 \left(12 M^3-20 r_0 M^2+17 r_0^2 M-4 r_0^3\right)\right) r_1-2 r_0 \left(Q^2+r_0 (r_0-2 M)\right)^2\right.\right.
        \\ \notag
        &\left.\left. \times \left(r_0 (5 r_0-2 M) Q^2-2 Q^4+r_0^2 \left(8 M^2-13 r_0 M+4 r_0^2\right)\right)\right]+\frac{2}{C Q} \left\{C r_0 \left(Q^2+r_0^2-2 Mr_0\right)\right.\right.
        \\ \notag
        &\left.\left.\times \left[2 (2 M-3 r_0) \left(Q^2-M r_0\right) r_0^q r_1 M^2-\left(Q^4+r_0 (3 r_0-4 M) Q^2+2 M (M-r_0) r_0^2\right) r_1^q M^2\right.\right.\right.
        \\ \notag
        &\left.\left.\left.-2 (2 M-3 r_0) r_0 \left(Q^2+r_0 (r_0-2 M)\right)^2 r_0^q+8 (M-r_0)^2 r_0^2 \left(Q^2+r_0 (r_0-2 M)\right) r_0^q\right]\right.\right.
        \\ \notag
        &\left.\left.-C r_0 \left(Q^2-M r_0\right) \left(Q^2+r_0 (r_0-2 M)\right) \left[8 (r_0^q r_1+r_0 (r_1^q-3 r_0^q)) M^3\right.\right.\right.
        \\ \notag
        &\left.\left.\left.+\left((8 r_0^q-5 r_1^q) Q^2+3 r_0 (4 r_0 r_0^q-2 r_1 r_0^q-r_0 r_1^q)\right) M^2+4 r_0 \left(3 Q^2+2 r_0^2\right) r_0^q M-4 Q^2 \left(Q^2+3 r_0^2\right) r_0^q\right]\right.\right.
        \\ \notag
        &\left.\left.+\frac{1}{\left(Q^2-M r_0\right) \mathcal{K}}\left[6 r_0^2 (r_1-2 r_0) M^4-r_0 \left((9 r_1-20 r_0) Q^2+r_0^2 (4 r_0+r_1)\right) M^3\right.\right.\right.
        \\ \notag
        &\left.\left.+2 \left(9 r_0^5+5 Q^2 r_0^3+2 Q^4 (r_1-2 r_0)\right) M^2-2 \left(5 r_0^6+9 Q^2 r_0^4+6 Q^4 r_0^2\right) M+2 r_0 \left(Q^2+r_0^2\right) \right.\right.
        \\ \notag
        &\left.\left.\left. \times \left(2 Q^4+r_0^2 Q^2+r_0^4\right)\right] \left[2 Q r_0 \left(11 Q^{10}+r_0 (41 r_0-87 M) Q^8+r_0^2 \left(255 M^2-209 r_0 M+29 r_0^2\right) Q^6\right.\right.\right.\right.
        \\ \notag
        &\left.\left.\left.\left.+r_0^3 \left(7 r_0^3-360 M^3+400 r_0 M^2-106 r_0^2 M\right) Q^4+M r_0^4 \left(252 M^3-348 r_0 M^2+135 r_0^2 M-17 r_0^3\right) Q^2\right.\right.\right.\right.
        \\ \notag
        &\left.\left.\left.\left.-M r_0^5 \left(72 M^4-120 r_0 M^3+64 r_0^2 M^2-14 r_0^3 M+r_0^4\right)\right) \mathcal{K}^{3/2}+C \left(M r_0-Q^2\right) r_0^q \left((2 r_0^2-3 Mr_0)\right.\right.\right.\right.
        \\ \notag
        &\left.\left.\left.\left.\times \left(Q^2+r_0 (r_0-2 M)\right)+\left(3 Q^2+r_0 (7 r_0-10 M)\right) \mathcal{K}\right)\right]+2 \left[4 r_0 (3 r_0-2 r_1) M^3\right.\right.\right.
        \\ \notag
        &\left.\left.\left.+\left((5 r_1-8 r_0) Q^2+r_0^2 (3 r_1-4 r_0)\right) M^2-2 \left(r_0^4+3 Q^2 r_0^2\right) M+4 Q^2 r_0 \left(Q^2+r_0^2\right)\right]\right.\right.
        \\ \notag
        &\left.\left. \times \left[C \left(Q^2-M r_0\right) \left(Q^2+r_0 (3 r_0-4 M)\right) r_0^q+Q r_0 \left(2 r_0 (64 M-29 r_0) Q^8-16 Q^{10}\right.\right.\right.\right.
        \\ \notag
        &\left.\left.\left.\left.+r_0^2 \left(313 r_0 M-386 M^2-48 r_0^2\right) Q^6+r_0^3 \left(564 M^3-631 r_0 M^2+184 r_0^2 M-14 r_0^3\right) Q^4\right.\right.\right.\right.
        \\ \notag
        &\left.\left.\left.\left.+M r_0^4 \left(-408 M^3+572 r_0 M^2-240 r_0^2 M+33 r_0^3\right) Q^2\right.\right.\right.\right.
        \\ \notag
        &\left.\left.\left.\left.+M r_0^5 \left(120 M^4-204 r_0 M^3+116 r_0^2 M^2-27 r_0^3 M+2 r_0^4\right)\right) \sqrt{\mathcal{K}}\right]\right.\right.
        \\ \notag
        &\left.\left.
        -2 \left[\left(Q^4+r_0 (3 r_0-4 M) Q^2+2 M (M-r_0) r_0^2\right) r_1 M^2+2 (M-r_0) r_0^2 \left(r_0^2-2 M r_0+Q^2\right)^2\right]\right.\right.
        \\ \notag
        &\left.\left.\times \left[3 C \left(r_0^2-2 M r_0+Q^2\right) r_0^q-2 Q r_0 \left(6 Q^8+8 r_0 (3 r_0-5 M) Q^6+r_0^2 \left(84 M^2-81 r_0 M+10 r_0^2\right) Q^4\right.\right.\right.\right.
        \\ 
        &\left.\left.\left.\left.-2 M r_0^3 \left(36 M^2-44 r_0 M+9 r_0^2\right) Q^2+M r_0^4 \left(24 M^3-36 r_0 M^2+12 r_0^2 M-r_0^3\right)\right) \sqrt{\mathcal{K}}\right]\right\}\right\}, 
    \end{align}
    \begin{align}
    \notag
        \xi_7&=2 E_1 M^2 Q \left(Q^2-M r_0\right) \left[6 Q^8+8 r_0 (3 r_0-5 M) Q^6+r_0^2 \left(84 M^2-81 r_0 M+10 r_0^2\right) Q^4\right.
        \\ \notag
        &\left.-2 M r_0^3 \left(36 M^2-44 r_0 M+9 r_0^2\right) Q^2+C+M r_0^4 \left(24 M^3-36 r_0 M^2+12 r_0^2 M-r_0^3\right)\right] \mathcal{K} r_0^3
        \\ \notag
        &-M^2 Q \left(r_0^2-2 M r_0+Q^2\right) \left\{2 E_1 r_0^2 \left(r_0^2-2 M r_0+Q^2\right) \left[5 Q^6+7 r_0 (r_0-3 M) Q^4\right.\right.
        \\ \notag
        &\left.\left.+M r_0^2 (27 M-14 r_0) Q^2-M r_0^3 \left(12 M^2-9 r_0 M+r_0^2\right)\right]-L_1 \sqrt{M r_0-Q^2} \left[6 Q^8+8 r_0 (3 r_0-5 M) Q^6\right.\right.
        \\ \notag
        &\left.\left.+r_0^2 \left(84 M^2-81 r_0 M+10 r_0^2\right) Q^4-2 M r_0^3 \left(36 M^2-44 r_0 M+9 r_0^2\right) Q^2+C\right.\right.
        \\ \notag
        &\left.\left.+M r_0^4 \left(24 M^3-36 r_0 M^2+12 r_0^2 M-r_0^3\right)\right]\right\} \mathcal{K} r_0+\left(Q^2-M r_0\right) \left\{C Q \left(5 Q^2+r_0 (3 r_0-8 M)\right) r_1 M^2\right.
        \\ \notag
        &\left.+C r_0 \left[\left(r_0^2-2 M r_0+Q^2\right) \left((E_1^q r_0-4 Q) M^2-4 Q r_0 M+4 Q \left(Q^2+r_0^2\right)\right)\right.\right.
        \\ \notag
        &\left.\left.-2 E_1 M^2 \mathcal{K} r_0^q\right]+2 Q \left[6 Q^8+8 r_0 (3 r_0-5 M) Q^6+r_0^2 \left(84 M^2-81 r_0 M+10 r_0^2\right) Q^4\right.\right.
        \\ \notag
        &\left.\left.-2 M r_0^3 \left(36 M^2-44 r_0 M+9 r_0^2\right) Q^2+M r_0^4 \left(24 M^3-36 r_0 M^2+12 r_0^2 M-r_0^3\right)\right] \left[\mathcal{K} r_1 M^2\right.\right.
        \\ \notag
        &\left.\left.+2 r_0 \left(Q^2-M^2-r_0 M+r_0^2\right) \left(r_0^2-2 M r_0+Q^2\right)\right]\right\} \sqrt{\mathcal{K}} r_0+\left(r_0^2-2 M r_0+Q^2\right)
        \\ \notag
        &\times \left\{2 C \left(M r_0-Q^2\right) \left(10 Q^2+3 r_0 (r_0-4 M)\right) r_0^q r_1 M^2-2 C r_0 \left(M r_0-Q^2\right) \mathcal{K} r_1^q M^2\right.
        \\ \notag
        &\left.+C L_1^q r_0 \left(r_0^2-2 M r_0+Q^2\right) \sqrt{M r_0 \mathcal{K}-Q^2 \mathcal{K}} M^2-4 Q r_0 \left(r_0^2-2 M r_0+Q^2\right) \left[5 Q^6+7 r_0 (r_0-3 M) Q^4\right.\right.
        \\ \notag
        &\left.\left.+M r_0^2 (27 M-14 r_0) Q^2-M r_0^3 \left(12 M^2-9 r_0 M+r_0^2\right)\right] \sqrt{\mathcal{K}} \left(r_1 \mathcal{K} M^2+2 r_0 \left(Q^2-M^2-r_0 M+r_0^2\right)\right.\right.
        \\ \notag
        &\left.\left.\times \left(r_0^2-2 M r_0+Q^2\right)\right)+2 C r_0^q \sqrt{M r_0-Q^2} \left[3 r_0 \left(4 r_0 \sqrt{M r_0-Q^2}+L_1 \sqrt{\mathcal{K}}\right) M^3\right.\right.
        \\ \notag
        &\left.\left.-\left(2 \left(4 r_0 \sqrt{M r_0-Q^2}+L_1 \sqrt{\mathcal{K}}\right) Q^2+r_0^2 \left(L_1 \sqrt{\mathcal{K}}-4 r_0 \sqrt{M r_0-Q^2}\right)\right) M^2\right.\right.
        \\
        &\left.\left.-6 r_0^2 \sqrt{M r_0-Q^2} \left(3 Q^2+r_0^2\right) M+8 Q^2 r_0 \left(Q^2+r_0^2\right) \sqrt{M r_0-Q^2}\right]\right\}.
    \end{align}
In the above expressions, $C$ is given in Eq.~\eqref{den} and for convenience we introduced $\mathcal{K}= r_0^2-3 M r_0+2 Q^2$.

\bibliography{JCAPBibliography.bib} 

\end{document}